\documentclass{article}
\usepackage{verbatim}
\usepackage{amssymb,pstricks,amsmath,amsthm}
\begin{document}
\newtheorem{cor}{Corollary}
\newtheorem{theorem}[cor]{Theorem}
\newtheorem{proposition}[cor]{Proposition}
\newtheorem{lemma}[cor]{Lemma}
\theoremstyle{definition}
\newtheorem{defi}[cor]{Definition}
\theoremstyle{remark}
\newtheorem{remark}[cor]{Remark}
\newtheorem{example}[cor]{Example}
\newcommand{\bX}{{\partial X}}
\newcommand{\cC}{\mathcal{C}}
\newcommand{\cDzu}{{{\mathcal D}_\Phi^{0,1}(X)}}
\newcommand{\cN}{{\mathcal N}}
\newcommand{\cz}{{\mathbb C}}
\newcommand{\cun}{\cC^{\infty}}
\newcommand{\dom}{\mathrm{dom}}
\newcommand{\ind}{\mathrm{index}}
\newcommand{\nz}{{\mathbb N}}
\newcommand{\pc}{{\Psi_\Phi(\rz^4)}}
\newcommand{\psus}{{\Psi_{\mathrm{sus}({}^\Phi\!T^*S^2)-\phi}(S^3)}}
\newcommand{\produs}{\mathrm{prod}}
\newcommand{\ptsX}{{{}^\Phi\!T^*X}}
\newcommand{\ptX}{{{}^\Phi\!TX}}
\newcommand{\px}{{\partial_x}}
\newcommand{\rz}{{\mathbb R}}
\newcommand{\sgn}{\mathrm{sign}}
\newcommand{\Sf}{\mathrm{sf}}
\newcommand{\Spec}{\mathrm{Spec}}
\newcommand{\tf}{\tilde{f}}
\newcommand{\tn}{\tilde{\nabla}}
\newcommand{\Tr}{\operatorname{Tr}}
\newcommand{\verti}{\mathrm{vert}}
\newcommand{\zz}{{\mathbb Z}}
\newcommand{\vf}{\varphi}
\newcommand{\be}{\begin{equation}}
\newcommand{\ee}{\end{equation}}

\title{$L^2$-index of the Dirac operator
of generalized Euclidean Taub-NUT metrics}
\author{Sergiu Moroianu 
\thanks{E-mail:~~~ moroianu@alum.mit.edu}\\
{\small \it Institutul de Matematic\u a al Academiei Rom\^ ane,}\\
{\small \it P. O. Box 1-764, RO-014700 Bucharest, Romania}
\and
Mihai Visinescu 
\thanks{E-mail:~~~ mvisin@theory.nipne.ro}\\
{\small \it Department of Theoretical Physics,}\\
{\small \it National Institute for Physics and Nuclear Engineering,}\\
{\small \it Magurele, P.O.Box MG-6, RO-077125 Bucharest, Romania}}
\date{}
\maketitle

\begin{abstract}
We compute the axial anomaly for the Taub-NUT metric on $\rz^4$. We show that
the axial anomaly for the generalized Taub-NUT metrics introduced  by
Iwai and Katayama is finite, although the Dirac operator is not Fredholm. 
We show that the essential spectrum of the Dirac operator is the whole 
real line.

Pacs: 04.62.+v
\end{abstract}

\section{Introduction}

The Taub-Newman-Unti-Tamburino (Taub-NUT) metrics were found by Taub 
\cite{Taub} and extended by Newman-Unti-Tamburino \cite{NUT}. The Euclidean 
Taub-NUT metric has lately attracted much attention in physics. Hawking 
\cite{SH} has suggested  that the Euclidean Taub-NUT metric might give 
rise to the gravitational analog of the Yang-Mills instanton.
This metric is the space part of the line element of the celebrated 
Kaluza-Klein monopole of Gross and Perry and Sorkin. On the other hand, 
in the long distance limit, neglecting radiation, the relative motion 
of two monopoles is described by the geodesics of this space \cite{AH}.
The Taub-NUT family of metrics is also involved in many other modern 
studies in physics like strings, membranes, etc.

From the symmetry viewpoint, the geodesic motion in Taub-NUT space 
admits a ``hidden" symmetry of the Kepler type. 
We mention that the following two generalization of the Killing vector 
equation have become of interest in physics:
\begin{enumerate}
\item A symmetric tensor field $K_{\mu_1...\mu_r}$  is called a 
St\" ackel-Killing (S-K) tensor of valence $r$ if and only if
\[K_{(\mu_1...\mu_r;\lambda)} = 0.\]
The usual Killing vectors correspond to valence $r=1$ while the hidden 
symmetries are encapsulated in S-K tensors of valence $r>1$.
\item A  tensor $f_{\mu_1...\mu_r}$ 
is called a Killing-Yano (K-Y) tensor of valence $r$ if it is totally 
anti-symmetric and it satisfies the equation
\[f_{\mu_1...(\mu_r;\lambda)} = 0.\]
The K-Y tensors play an important role in models for relativistic 
spin-$\frac{1}{2}$ particles having in mind that they produce first-order 
differential operators of the Dirac-type which anticommute with the 
standard Dirac one \cite{CL}.
\end{enumerate}

The family of Taub-NUT metrics with their plentiful symmetries provides
an excellent background to investigate the classical and quantum 
conserved quantities on curved spaces. 
In the Taub-NUT geometry there are four K-Y tensors. 
Three of these are complex structures realizing the quaternion algebra 
and the Taub-NUT manifold is hyper-K\" ahler \cite{GR}. In addition to 
these three vector-like K-Y tensors, there is a scalar one 
which has a non-vanishing field strength and which exists by virtue of the 
metric being type D.

For the geodesic motions in the Taub-NUT space, the conserved vector 
analogous to the Runge-Lenz vector of the Kepler type problem  is 
quadratic in 4-velocities, and its components are 
S-K tensors which can be expressed as symmetrized 
products of  K-Y tensors \cite{GR,VV1}.

To the hidden symmetry encapsulated into S-K tensor $k_{\mu\nu}$, the 
corresponding quantum operator is
\[{\cal K} = D_\mu k^{\mu\nu} D_\nu\]
where $D_\mu$ is the covariant differential operator on the 
curved manifold.  It commutes with the scalar Laplacian
\[{\cal H} = D_\mu D^\mu\]
if the space is Ricci flat. That is the case for the standard 
Taub-NUT space which is hyper-K\" ahler. Moreover, the commutator 
$ [{\cal H}, {\cal K}] $ vanishes even for Ricci non-flat spaces if the 
S-K tensor $k_{\mu\nu}$ can be expressed as a symmetrized product of 
K-Y tensors \cite{CL}.

Iwai and Katayama \cite{IK1,IK2,IK3} generalized the Taub-NUT 
metrics in the following way. Suppose that a metric $\bar g$ on 
an open interval $U$ in $(0, + \infty $) and a family of Berger 
metrics $\hat g (r)$ on $S^3$ indexed by $U$ are given, where a family 
of Berger metric is by definition a right invariant metric on $S^3 = Sp(1)$  
which is further left $U(1)$ invariant. Then the twisted product $g = 
\bar g + \hat g (r)$ on the annulus $U\times S^3 \subset \rz^4\setminus\{0\}$ 
is called a generalized Taub-NUT metric \cite{YM}.
In what follows we shall restrict to such generalizations which
admit the same Kepler-type symmetry as the standard Taub-NUT metric.
These metrics are defined on $\rz^4\setminus\{0\}$ by the line element
\[
\begin{split}
{ds_K}^2&=g_{\mu\nu}(x)dx^{\mu}dx^{\nu}\\
&=f(r)(dr^2+r^2d\theta^2+r^2\sin^2\theta\, d\varphi^2)
+g(r)(d\chi+\cos\theta\, d\varphi)^2
\end{split}
\]
where the angle variables $(\theta,\varphi,\chi)$ parametrize the sphere
$S^3$ with $ 0\leq\theta<\pi, 0\leq\varphi<2\pi, 
0\leq\chi<4\pi$, while the functions 
\begin{align*}
f(r) = \frac{ a + b r}{r},&& g(r) = 
\frac{ a r + b r^2}{1 + c r + d r^2}.
\end{align*}
depend on the arbitrary real constants $a,\,b,\,c$ and $d$. The singularity at 
$r=0$ disappears by the change of variables $r=y^2$, hence ${ds_K}^2$ is
a complete metric on $\rz^4$. For positive definiteness, we assume
that $a,b,d>0$, and $c>-2\sqrt{d}$.  
If one takes the constants 
\begin{align*}
c=\frac{2 b}{a},&& d = \frac{b^2}{a^2}
\end{align*} 
the generalized Taub-NUT metric becomes the original
Euclidean Taub-NUT metric up to a constant factor.

The necessary condition that a S-K tensor of valence two be written 
as the square of a K-Y tensor is that it has at the most two distinct 
eigenvalues \cite{MV}. In the case of the generalized Taub-NUT spaces 
the S-K tensors involved in the Runge-Lenz vector cannot be expressed 
as a product of K-Y tensors.
The non-existence of the K-Y tensors on generalized Taub-NUT metrics 
leads to gravitational quantum anomalies proportional to a contraction 
of the S-K tensor with the Ricci tensor \cite{tanug}.

In our previous paper \cite{tanug} we computed the axial quantum anomaly,
interpreted as the index of the Dirac operator of these metrics,
on annular domains and on disks, with the non-local
Atiyah-Patodi-Singer boundary condition. We found that the index is 
a number-theoretic quantity which depends on the coefficients of the metric.
In particular, our formula shows that this index vanishes on balls of 
sufficiently large radius, but can be nonzero for some values of the parameters
$c,d$ and of the radius.

We also examined the Dirac operator on the complete Euclidean space with 
respect to this metric, acting in the Hilbert space of square-integrable 
spinors. We found  that this operator is not Fredholm, hence even the 
existence of a finite index is not granted.

We mentioned in \cite{tanug} some open problems in 
connection with unbounded domains. The present work brings new 
results in this direction. First we 
show that the Dirac operator on $\rz^4$ with respect to the 
standard Taub-NUT metric does not have $L^2$ harmonic spinors. This 
follows rather easily from the Lichnerowicz formula, since the 
standard Taub-NUT metric has vanishing scalar curvature. In particular, the 
index vanishes.

Entirely different techniques are needed for the generalized Taub-NUT metrics, 
since they are no longer scalar-flat.
We first note that the essential spectrum of the associated
Dirac operator is $\rz$,
and we describe its domain. This is a direct application of the work done in
\cite{tanug} and of the theory of $\Phi$-pseudodifferential calculus
developed in \cite{mame99}. Next 
we show that the dimension of the kernel is finite. This is by no means
easy. The standard way of getting such a finiteness result
is proving that the operator is Fredholm on a larger $L^2$ space. This 
approach works for
$b$- or cusp operators via a conjugation argument (see \cite{melaps})
but it fails for $\Phi$-operators when the 
dimension of the base is greater than $0$, as is the case here.

Nevertheless, by applying the main result of \cite{boris}, 
we manage to show that the 
dimension of the kernel is finite. We must still leave open the question of 
computing the index. We conjecture that it equals $0$ and hence, unlike 
on annular domain or balls, the axial anomaly is never present. Our 
guess is motivated by heuristically increasing the radius of a ball to infinity,
and arguing that by \cite{tanug}, the index stabilizes at $0$ for large radii.
Such an argument is of course incomplete, and even dangerous in the light
of the fact that the Dirac operator is not Fredholm.

\section{The axial anomaly}

Let $D$ denote the Dirac operator for the metric ${ds_K}^2$, acting as an 
unbounded operator
in $L^2(\rz^4,\Sigma_4,{ds_K}^2)$ with initial domain $\cun_c$. The generalized 
Taub-NUT metric is complete and smooth on $\rz^4$, hence 
the Dirac operator is essentially self-adjoint.

We proved in \cite{tanug} that $D$ is not Fredholm. This implies that even a 
small perturbation could in principle change the index of the chiral part of 
$D$. Moreover, it is not at all clear that the kernel of $D$ is 
finite-dimensional! This makes the computation of the $L^2$ 
index rather delicate.

\begin{theorem}
For the standard Taub-NUT metric on $\rz^4$ 
the Dirac operator does not have $L^2$ solutions.
\end{theorem}
\begin{proof}
Recall that the standard Taub-NUT metric is hyper-K\"ahler, hence its scalar 
curvature $\kappa$ vanishes.

By the Lichnerowicz formula, 
\[D^2=\nabla^*\nabla+\frac\kappa{4}=\nabla^*\nabla.\]

Let $\phi\in L^2$ be a solution of $D$ in the sense of distributions. 
Then, again in distributions, $\nabla^*\nabla\phi=0$.  The operator
$\nabla^*\nabla$ is essentially self-adjoint with domain 
$\cun_c(\rz^4,\Sigma_4)$, which implies that its kernel equals the kernel
of $\nabla$. Hence $\nabla\phi=0$. 
Now a parallel spinor has constant pointwise norm, hence
it cannot be in $L^2$ unless it is $0$, because the volume of the metric
${ds_K}^2$ is infinite. Therefore $\phi=0$.
\end{proof}

We turn now to the generalized Taub-NUT metrics. We refer to \cite{tanug}
for previous results on the quantum anomalies of these metrics on annular 
domains and on balls. 

We have noticed in \cite{tanug} that ${ds_K}^2$ belongs to the class of
fibered cusp metrics from \cite{mame99}. Moreover, its Dirac operator $D$
is elliptic but not fully elliptic in this calculus. 
\begin{proposition}
Every elliptic symmetric $\Phi$-operator $A$ of order $a\geq 0$ with initial 
domain $\cun_c$ is essentially self-adjoint, and its domain is
the fibered-cusp Sobolev space $H^a_\Phi$.
\end{proposition}
\begin{proof}
Since $A\in\Psi_\Phi^a$
is elliptic, there exists $G\in \Psi_\Phi^{-a}$ an inverse of $A$
modulo $\Psi_\Phi^{-\infty}$. Thus
\begin{align*}
AG=1-R_1,&&GA=1-R_2
\end{align*}
where $R_1,R_2$ belong to $\Psi_\Phi^{-\infty}$. Note that $R_1,R_2$ do 
not have to be compact operators.
Recall from \cite{mame99}
the mapping properties of $\Phi$ operators: as in the closed manifold case,
an operator $A$ of order $a$ maps $H_\Phi^k$ into $H^{k-a}_\Phi$ for all real 
$k$, and moreover for $a\geq 0$, $H^a$ is contained in the domain of the 
closure of $A$ with initial domain $\cun_c$. 
If $\phi\in L^2$ is in the domain of the adjoint of $A$, i.e., 
$A\phi\in L^2$ in the sense of distributions, then
\[A\phi\in H^0_\Phi=L^2 \Rightarrow GA\phi\in H^a_\Phi.\]
This means $\phi-R_2\phi\in H^a_\Phi$. Now 
\[\phi\in H^0_\Phi \Rightarrow R_2\phi\in H^\infty_\Phi.\]
This implies that $\phi$ belongs to $H^a_\Phi$. Hence
\[H^a_\Phi\subset \dom(\overline{A})\subset\dom(A^*)\subset H^a_\Phi\]
which ends the proof. 
\end{proof}

From \cite{mame99}, $D$ is Fredholm from its domain $H^1_\Phi$ to $L^2$ 
if and only if
it is fully elliptic. Thus let us compute its normal operator. Outside
$0\in\rz^4$ we set $x=1/r$. Let
\begin{align*}
\alpha(x):=\frac{1}{\sqrt{ax+b}},&&\beta(x):=\sqrt{x^2+cx+d}.
\end{align*}
Let $I,J,K$ denote the vector fields on $S^3$ corresponding to the 
infinitesimal action of quaternion multiplication by the unit vectors
$i,j,k$. We trivialize the tangent bundle to 
$\rz^4\setminus\{0\}\simeq (0,\infty)\times S^3$ using the orthonormal frame
\begin{align*}
V_0=\alpha(x)x^2\px,&& V_1=\alpha(x)\beta(x)I/2,&&
V_3=\alpha(x)xJ/2,&&V_4=\alpha(x)xK/2.
\end{align*}
Denote by $c^j$ the Clifford multiplication with the vector $V_j$.
Since $\rz^4\setminus\{0\}$ is simply connected, there exists a lift
of this frame to the spin bundle. A long but 
straightforward computation shows
that in the trivialization of the spinor bundle given by this lift,
the Dirac operator equals
\[\begin{split}
D &=c^0\left(V_0-\frac{x^2}{2\beta(x)}(\alpha\beta)'-x(x\alpha)'\right)
+c^1\left(V_1-\frac{\alpha\beta}{2}c^2c^3\right)\\
&\quad +c^2 V_2+c^3 V_3+\frac{x^2\alpha}{4\beta}c^1c^2c^3.
\end{split}\]

We assume for simplicity that $b=1$. We can always reduce 
ourselves to this case by a scalar conformal change of the metric.

The normal operator is obtained in two steps (see \cite{mame99}). 
We first formally replace
$x^2\px$ with $i\xi$, $\xi\in\rz$, and $xJ/2,xK/2$ with $\tau_2,\tau_3$
where $\tau_2,\tau_3\in \rz$ are global coordinates on the vector bundle
$\phi^* TS^2$ over $S^3$ (note that $TS^2$ is not trivial, but its 
pull-back to $S^3$ through the Hopf fibration is). The second step consists
in freezing the coefficients at $x=0$. Thus for
$\xi\in\rz,\tau\in \phi^* TS^2$,
\[\cN(D)(\xi,\tau)=i\xi c^0+ic(\tau)+
D_\verti\]
where 
\[D_\verti=c^1\frac{\sqrt{d}}{2}\left(I-c^2c^3\right)\]
is a family of differential operators on the fibers
of the Hopf fibration $S^3\to S^2$ (recall that $I$ is a vector field
with closed trajectories of length $2\pi$). We have observed in
\cite{tanug} that $D_\verti$ is not invertible. Indeed,
$c^2c^3$ is skew-adjoint of square $-1$ and hence $\exp(2\pi c^2c^3)=1$. 
This shows that $\ker(D_\verti)$ is made of spinors $\psi$ satisfying 
\[\psi(e^{it} p)=e^{tc^2c^3}\psi(p)\]
(the multiplication is in the sense of quaternions). 
The space of such spinors on the fiber over each point in $S^2$ has 
complex dimension equal to $\dim(\Sigma(4))=4$.

\begin{theorem}
The essential spectrum of $D$ is $\rz$.
\end{theorem}
\begin{proof}
Equivalently, since $D$ is self-adjoint, we show that for all $\lambda\in\rz$, 
$D-\lambda$ is not Fredholm. By the discussion above,
$D-\lambda$ is Fredholm if and only if it is fully elliptic, i.e.,
if and only if $\cN(D-\lambda)(\xi,\tau)$ is invertible as a family
of operators on the fibers of the Hopf fibration for all $\xi,\tau$.
Fix a point $p$ in $S^2$. Then on the kernel of $D_\verti$ on the fiber 
over $p$,
\[\cN(D-\lambda)(\xi,\tau)=i\xi c^0+ic(\tau)-\lambda.\]
Set $\tau=0$; for $\xi=\lambda$, the spectrum of the matrix 
$i\xi c^0$
is $\{\pm\lambda\}$, so $i\xi c^0-\lambda$ cannot be invertible for all
real $\xi$. 
\end{proof}

\begin{remark}
Note that $D$ is not Fredholm on any weighted $L^2$ space 
$e^{\frac{\gamma}{x}}L^2$. This is because the conjugate
$e^{-\frac{\gamma}{x}}De^{\frac{\gamma}{x}}$ acting in $L^2(\rz^4,{ds_K}^2)$
has normal operator
\[\cN(e^{-\frac{\gamma}{x}}De^{\frac{\gamma}{x}})(\xi,\tau)=c^0(i\xi-\gamma)+
ic(\tau)+D_\verti.\]
This operator vanishes on a spinor which fiberwise is in 
the kernel of $D_\verti$, for $\xi=0$ and for a vector $\tau$ with
$|\tau|^2=\gamma^2$. 
\end{remark}
Nevertheless, we can prove the following finiteness result:

\begin{theorem}
The $L^2$ kernel of $D$ has finite dimension.
\end{theorem}
\begin{proof}
Although the dimension of $D_\verti$ is not zero, it is at least constant
when the base point in $S^2$ varies. Let $h:[0,\infty)\to[1,\infty)$ be a smooth
function which equals $r(a+br)$ for large $r$. Set
\[g_d:=h^{-1}{ds_K}^2.\]
This is a conformally equivalent metric which falls into the class of 
$d$-metrics studied by Vaillant \cite{boris}. Indeed, at infinity, in the 
variable $x=1/r$,
\[g_d=\frac{dx^2}{x^2}+g_H+x^2\frac{g_V}{x^2+cx+d^2}\]
where $g_H$ is a metric pulled back from the base, and $g_V$ is a family of 
metrics on the fibers of the Hopf fibration, both constant in $x$. 
This is an exact $d$-metric with 
constant dimensional kernel of the ``vertical" Dirac operator,
thus by \cite[Chapter 3]{boris}, its Dirac operator  $D_d$ has 
finite-dimensional kernel in $L^2(\rz^4,\Sigma(4),dg_d)$.

We apply now the conformal change formula for the Dirac operator 
\[D=h^{-5/4}D_d h^{3/4}\]
(see e.g.,
\cite[Appendix A.2]{boris}). If $\phi\in L^2({ds_K}^2)$ is in the null-space 
of $D$ then $h^{3/4}\phi\in\ker(D_d)$. Moreover, 
\[\|h^{3/4}\phi\|^2_{L^2(g_d)}=\int_{\rz^4}h^{-1/2}|\phi|^2 {ds_K}^2\]
is finite since $h^{-1}$ is bounded.
We obtained an injection of $\ker(D)$ into the finite-dimensional space
$\ker(D_d)$.
\end{proof}

\subsection*{Acknowledgments}
MV would like to acknowledge Emilio Elizalde, Sergei Odintsov and the 
Organizing Committee of the Seventh Workshop QFEXT'05 for the 
hospitality and financial support.
SM has been partially supported by the
contract MERG-CT-2004-006375
funded by the European Commission, and by a CERES contract (2004), Romania.
MV has been partially supported by a grant MEC-CNCSIS, Romania.

\bibliographystyle{amsplain}

\end{document}